# Persistent antiferromagnetic order in heavily overdoped $Ca_{1-x}La_xFeAs_2$

*Edoardo Martino[a,d], Maja D. Bachmann[d], Lidia Rossi[a], Kimberly A. Modic[d], Ivica Zivkovic[b], Henrik M. Rønnow[b],*

*Philip J. W. Moll[c,d], Ana Akrap[e], László Forró[a] and Sergiy Katrych[a]*

[a]*Laboratory of Physics of Complex Matter, Institute of Physics, École Polytechnique Fédérale de Lausanne (EPFL), CH-1015 Lausanne, Switzerland*
[b]*Laboratory for Quantum Magnetism, Institute of Physics, Ecole Polytechnique Féderale de Lausanne (EPFL), CH-1015 Lausanne, Switzerland*
[c]*Laboratory for Quantum Materials, Institute of Materials, École Polytechnique Fédérale de Lausanne (EPFL), CH-1015 Lausanne, Switzerland*
[d]*Max-Planck-Institute for Chemical Physics of Solids, D-01187 Dresden, Germany*
[e]*Department of Physics, University of Fribourg, CH-1700 Fribourg, Switzerland*

**Abstract**

In the $Ca_{1-x}La_xFeAs_2$ (112) family of pnictide superconductors, we have investigated a highly overdoped composition (x = 0.56), prepared by high-pressure, high-temperature synthesis. Magnetic measurements show an antiferromagnetic transition at $T_N$ = 120 K, well above the one at lower doping (0.15 < x < 0.27). Below the onset of long-range magnetic order at $T_N$, the electrical resistivity is strongly reduced and is dominated by electron-electron interactions, as evident from its temperature dependence. The Seebeck coefficient shows a clear metallic behavior as in narrow band conductors. The temperature dependence of the Hall coefficient and the violation of Kohler's rule agree with the multiband character of the material. No superconductivity was observed down to 1.8 K. The success of the high-pressure synthesis encourages further investigations of the so far only partially explored phase diagram in this family of Iron-based high temperature superconductors.

*Introduction*

Since the discovery of high temperature superconductivity in the pnictides [1], numerous compounds have been synthesized, notably the families: $(Ba,K)Fe_2As_2$ (122) [2], FeSe (11) [3], LiFeAs (111) [4], $LaFeAsO_{1-x}(H,F)_x$ (1111) [5], and most recently, the $Ca_{1-x}La_xFeAs_2$ (112) [6]. The 112-family distinguishes itself from the others by having a metallic block-layer between the FeAs planes. Such metallic layer, made of As zig-zag chains (Fig. 1. A) [6], adds one hole-like pocket and four bands with a Dirac-cone-like dispersion to the band structure [7]. It is claimed that the possible coexistence of Dirac electrons and superconductivity in the FeAs layers, can potentially lead to the observation of Majorana fermions [8]. The other striking peculiarity of the $Ca_{1-x}La_xFeAs_2$ 112-family becomes apparent in its temperature-doping phase diagram. It is only partially explored, as only the doping range 0.15 < x < 0.27 has been synthesised to date, marked as the solid background area in Fig. 1B. It is known that in $Ca_{1-x}La_xFeAs_2$ superconductivity appears with its highest observed critical temperature of $T_C$ = 35 K for the lowest reported doping level (x = 0.15). For higher doping, $T_C$ decreases until its full suppression at x = 0.23. Collinear antiferromagnetic order (AFM) coexists microscopically with superconductivity [9,10] in the doping range of 0.15 < x < 0.23. The Néel temperature ($T_N$) constantly increases with doping, even beyond the suppression of the superconducting phase [10].

The magnetic order goes hand-in-hand with a structural phase transition, from monoclinic to triclinic symmetry, which is driven by magneto-elastic coupling [9], a common feature in the pnictides [11]. Another commonality among the pnictide families is the manifestation of a kink in the temperature dependence of the electrical resistivity at the transition [11]. This is due to reduced scattering from magnetic fluctuations below $T_N$ [9].

For a better understanding of this family, and the pnictide superconductors in general, it is important to explore the phase diagram in a broad doping range, especially on the overdoped side, above x = 0.27. It would be highly informative to check if there is a second superconducting dome, as in the 1111-family [12], and to see how the antiferromagnetic order evolves with increasing carrier density.

In this quest, we used high-pressure synthesis to make heavily over-doped single crystals of $Ca_{1-x}La_xFeAs_2$. Structural refinement of single crystal X-ray diffraction (XRD) data yields x = 0.56. Magnetic susceptibility and XRD measurements suggest that the antiferromagnetic phase transition occurs around 120 K. Transport measurements show bad metal behaviour, with strong interband scattering, and no superconducting transition was observed down to 1.8 K.

Our measurements add an important point to the phase diagram of the 112-family (Fig. 1. B), suggesting the persistence of the AFM order even at higher doping levels. Compared to other pnictides families, in 112-family the AFM order persists over a considerably wider doping range [13,14].

**High-pressure synthesis**

High-pressure, high-temperature synthesis proved to be a valuable approach to obtain novel materials and to stabilize phases and compositions that are not accessible by other crystal growth techniques [15,16,17]. Another advantage of the high-pressure synthesis is that the chemical reaction happens in a strongly confined space. If the reaction were prepared inside a quartz ampulla for example, because As has a high vapour pressure, the developing overpressure at high temperatures could lead to an explosion.

For the high-pressure synthesis, 0.6 g mixture of starting components was used. As, Ca, CaO, FeAs, La, respectively with the molar ratio of 1:0.6:1:2:0.4 were placed in a BN crucible with 0.4 g of NaCl flux. The reactants where pressurized at 3 GPa and then heated to 1430 °C within 2 h. The temperature was maintained for 5 h, then slowly (7.5 °C/h) cooled down to 1000 °C and annealed at this temperature for 2 h. Afterwards, the system was rapidly cooled down to room temperature by switching off the power supply. The NaCl flux was dissolved in distilled water. Plate-like single crystals, of $Ca_{1-x}La_xFeAs_2$, with approximate dimensions of 10 x 120 x 180 µm$^3$ were selected for further investigations. In addition, a large number of small crystallites of FeAs phase were formed, as well.

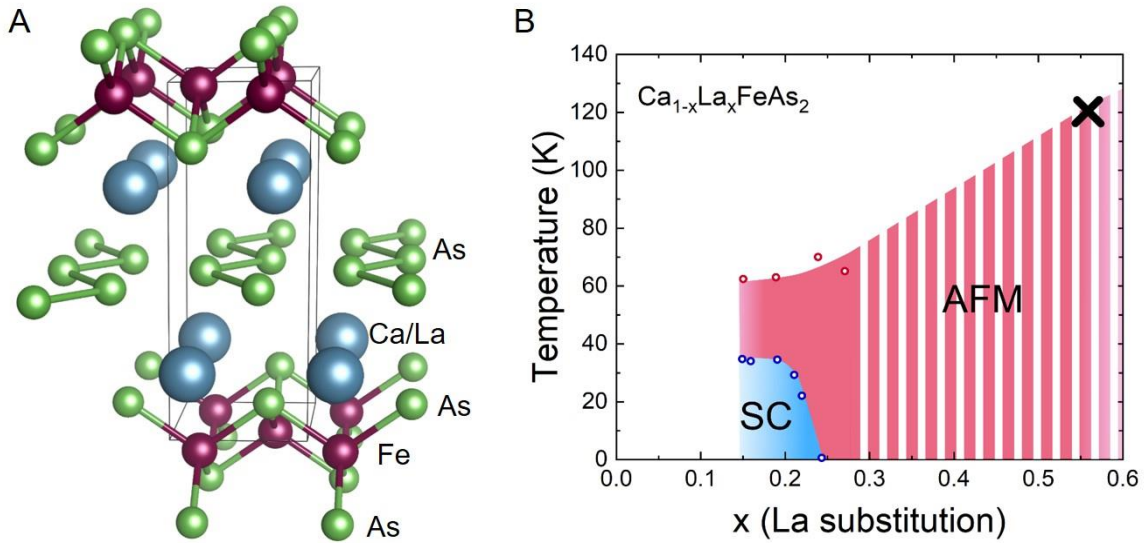

Figure 1. A) Sketch of the $Ca_xLa_{1-x}FeAs_2$ crystal structure. B) The proposed electronic phase diagram for $Ca_xLa_{1-x}FeAs_2$ as a function of the substitutional doping. Data from previously reported synthesis are shown as open circles in the region with solid background, for 0.15 < x < 0.27 [10]. The black cross is the result of the present study, showing the antiferromagnetic transition temperature (120 K) for x = 0.56. Striped area is an extrapolation of the AFM phase.

*Single crystal XRD*

The two majority phases of FeAs and $Ca_{1-x}La_xFeAs_2$ were identified by Rietveld analysis of the powder X-ray diffraction (XRD) spectra, with mass ratios of 86 % and 14 %, respectively. Single crystal X-ray diffraction was performed on selected crystals of $Ca_{1-x}La_xFeAs_2$. The diffraction pattern was collected using a SuperNova (dual Source) four-circle diffractometer (Agilent Technologies, USA) equipped with a CCD detector. Data reduction and analytical absorption correction were made using the CrysAlis[Pro] software package [18]. The crystal structure was refined using known lattice parameters from [6] and the SCHELXL software package. The refinement of the spectra indicates a composition of x = 0.56(1). On the very same crystal, energy-dispersive X-ray spectroscopy (EDX), gave x = 0.57(2). For the many different crystals synthesised, EDX gave a slight variation in the composition in the 0.55 < x < 0.6 range. The main quantities deduced from single crystal X-ray diffraction are shown in Table 1.

**Table 1. Results of the structural refinement of $Ca_{0.44}La_{0.56}FeAs_2$.**

| | | |
|---|---|---|
| Crystal size | $10 \times 119 \times 180\ \mu m^3$ | |
| Empirical formula | $Ca_{0.44}La_{0.56}FeAs_2$ | |
| Temperature | 293(2) K | |
| Crystal symmetry | monoclinic | |
| Space group | $P2_1$ | |
| Unit cell dimensions | $a = 3.9463(7)$ Å | $\alpha = 90°$. |
| | $b = 3.9247(5)$ Å | $\beta = 90.792(15)°$. |
| | $c = 10.4652(18)$ Å | $\gamma = 90°$. |
| Volume | $162.07(5)$ Å$^3$ | |
| Density (calculated) | $6.180\ Mg/m^3$ | |
| Refinement method | Full-matrix least-squares on $F^2$ | |
| Goodness-of-fit on $F^2$ | 1.081 | |
| Final R indices [I > 2sigma(I)] | $R_1 = 0.0743$, $wR_2 = 0.1956$ | |
| R indices (all data) | $R_1 = 0.0809$, $wR_2 = 0.2018$ | |

## *The antiferromagnetic state*

The magnetic state of overdoped $Ca_{1-x}La_xFeAs_2$ was determined by the temperature-dependent magnetic susceptibility (Fig 2. A) which was measured in a MPMS SQUID. The data were collected in field cooling of 1 T. To enhance the signal, 14 mg of selected reaction products were introduced in a gelatine capsule for measurement, which contained randomly oriented crystals of $Ca_xLa_{1-x}FeAs_2$ and FeAs as impurity phase. Two kinks are present in the susceptibility measurement, one around 77 K and the other around 120 K. The kink at 77 K is associated with the magnetic transition present in FeAs [19]. The broad kink at higher temperature is compatible with the antiferromagnetic transition in $Ca_xLa_{1-x}FeAs_2$. Previously reported susceptibility data on single crystals show a different shape of the kink for different field orientations [9]. Since the sample we use for magnetic measurements is a collection of differently oriented crystallites, such anisotropic response is averaged, resulting in a smeared and less sharp kink. The small variation in composition in the range from x = 0.55 to x = 0.6 also contributes to the observed magnetization curve.

Another signature of the AFM state, observed in a previous investigation of $Ca_xLa_{1-x}FeAs_2$, is the lowering of the crystal symmetry from monoclinic to triclinic [20], via a magneto-elastic coupling. In order to verify this, around 20 single crystals of $Ca_{1-x}La_xFeAs_2$ were assembled in a glass capillary for temperature-dependent XRD measurements, performed with synchrotron radiation at the BM01 station of the Swiss-Norwegian Beam Lines (ESRF, Grenoble, France). The data were collected with Pilatus@SNBL diffractometer, and processed with SNBL ToolBox software [22]. The results in Fig. 2. B, show the evolution of the *220* Bragg peak with temperature. At room temperature, a single resolution-limited peak is observed at 2Θ = 29.77°. At lower temperatures, the peak shifts to higher 2Θ values due to the contraction of the unit cell. When the temperature is brought below $T_N$, the peak splits into two overlapping peaks, indicating that the symmetry is lowered (Fig. 2. B) from monoclinic to triclinic.

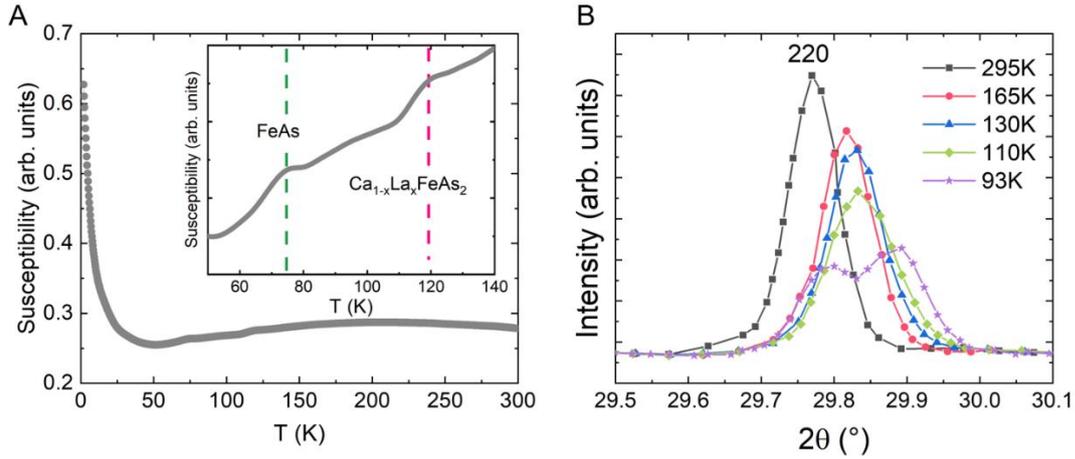

Figure 2. A) The temperature dependence of the magnetic susceptibility of a mixture of FeAs and $Ca_{1-x}La_xFeAs_2$ compositions. The inset shows an enlarged view of magnetic transitions of FeAs at 77 K (green) and $Ca_{1-x}La_xFeAs_2$ at ∼ 120 K (pink). B) The temperature dependence of the 220 diffraction peak (measured with λ = 0.71446 Å) shows a splitting which occurs between 130 K and 93 K as a consequence of the monoclinic to triclinic phase transition, provoked by the magneto-elastic coupling in the AFM state.

*Transport properties*

Electrical resistivity was measured on single crystals of well-defined composition. The phase purity and doping level were characterized by single crystal XRD and energy-dispersive X-ray spectroscopy (EDX). Using a focused ion beam (FIB), the single crystal was micro-fabricated into a Hall-bar geometry of size 110 x 11 x 6 µm$^3$ [23]. In this way the current path was defined along the *ab*-plane for in-plane resistivity (ρ), magnetoresistance and Hall coefficient ($R_{Hall}$) measurements. The temperature dependence of the in-plane electrical resistivity is shown in Fig 3 A. The precise geometry of the micro-fabricated sample enabled to measure the resistivity with high accuracy. Our sample shows a lower resistivity compared to the reported results for a lower doping level (x = 0.27) [9]. This could be attributed to the increased carrier density accompanying the overdoping, as observed in other pnictides [24]. On cooling, ρ(T) shows a weak temperature dependence down to 120 K, followed by a kink at the paramagnetic-AFM transition temperature. Below 120 K, the behaviour is more metallic but the residual resistivity ($ρ_0$) is relatively high. At this point, the origin of $ρ_0$ is unclear. One possibility is that it comes from static disorder introduced during the high-pressure synthesis.

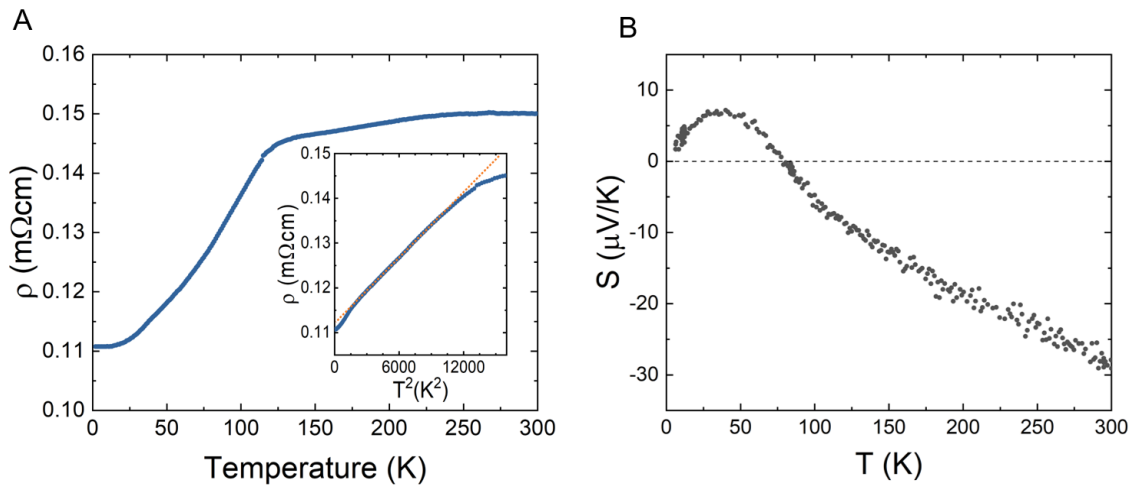

Figure 3. A) Temperature dependence of the in-plane electrical resistivity, measured on a focused ion beam micro-fabricated single crystal. Inset: The low temperature resistivity (T < 90 K) plotted as a function of T$^2$. The dotted orange line is a guide for the eyes. B) Seebeck coefficient versus temperature measured on a pellet sample of the overdoped Ca$_x$La$_{1-x}$FeAs$_2$ (0.55 < x < 0.6).

The temperature dependence of ρ in a broad range (between 35 K and 90 K) shows Fermi liquid behaviour that is characterized by the T$^2$ dependence: $\rho(T) = \rho_0 + AT^2$ (with ρ$_0$ = 0.11 mΩcm and $A$ = 0.0024 μΩcm/K$^2$) (inset to Fig 3. A). For lower temperatures (T < 35 K) the resistivity varies as T$^3$. Typical low temperature resistivity in metals scales as T$^5$ due to scattering between electrons and long-wavelength phonons, as described by the Bloch–Grüneisen relation. But for a Fermi-liquid, such a crossover in the power law can yield an exponent of 3 like we observe [25]. It is challenging to understand ρ(T) above the magnetic transition (T > T$_N$), where it shows a very weak temperature dependence. This bad metal character contrasts with the Seebeck coefficient (Fig 3. B) which displays a clear metallic dependence in the same temperature range. We propose that this behaviour originates from the increasing interband scattering with decreasing temperatures, reducing the resistivity temperature dependence, which opposes the effect of electron-phonon scattering. A similar mechanism was reported for other pnictides where multiple bands at the chemical potential favour strong interband scattering [26].

It has to be mentioned that the Mooij correlation is another possible source of the observed flat resistivity above T > T$_N$ [27]. In the case of a high concentration of static defects, marked by the elevated ρ$_0$, electronic interferences between the scattering events tend to weakly localize electrons. After a threshold value ρ$_{TH}$, even the dρ/dT changes to a negative value. In Mooij's original picture, ρ$_{TH}$ was equal to 0.2 mΩcm (close to the value in Fig 3. A, 0.15 mΩcm), but later on Tsuei [28] showed that ρ$_{TH}$ is a material dependent quantity. In our case, ρ$_0$ is indeed high, however, if localization effects dominate, one would not expect a major change in ρ(T) at T$_N$. Due to the strong change in the slope below T$_N$, we believe that the multiband scattering dominates the high temperature resistivity.

The temperature dependence of the Seebeck coefficient (S) (Fig. 3. B) may provide a further insight into the electronic properties of the material. In multiband systems, the Seebeck coefficient is given by $S = \frac{\sum_i \sigma_i S_i}{\sum_i \sigma_i}$, where $\sigma_i$ is the i$^{th}$-band conductivity and $S_i$ is its thermoelectric contribution, positive for holes and negative for electrons. The Seebeck coefficient is an open-circuit measurement, so it is not affected by grain boundaries or static defects. But in order to obtain a precise measurement one needs a well-defined linear temperature gradient on the sample, hence a sample much longer than the typical 100 μm size. For this reason, many small size single crystals were compressed into a pellet, from which a millimeter long bar was shaped, and used for the measurement.

The negative Seebeck coefficient at high temperatures suggests a dominant contribution of electrons, in agreement with the Hall coefficient (see Fig 4. A) and previous reports for samples with low doping [20]. The high temperature part is linear, and by using the Mott's formula: $S(T) = \frac{\pi^2}{3}\left(\frac{k_B}{e}\right)\left(\frac{T}{T_F}\right)$, we extract an effective Fermi temperature (T$_F$) of 3080 K (0.27 eV). The T$_F$ value and the A coefficient extracted from the resistivity temperature dependence fall on the same trend as a number of Fermi liquids, probing the relationship between the electron-electron scattering rate and Fermi energy [21]. This magnitude of T$_F$ suggests a relatively narrow conduction band. Just below the transition temperature, S changes sign, as R$_H$ increases, revealing an increased contribution of the hole bands to the overall conductivity. The additional hole pocket induced by the As-chains could be responsible for these observations.

Below T$_N$, S(T) changes slope, as it is defined by the energy dependence of the scattering rate, which is affected by the magnetic order and the Fermi surface reconstruction. Below this temperature, the relative contributions of the electron and hole bands change, and S changes sign. This is yet another manifestation of the multiband character of the material. In the T → 0 K limit, S goes to 0, as required by thermodynamics [29].

*Magneto-transport properties*

The Hall coefficient (R$_H$) (Fig 4. A) was measured on the same sample as the electrical resistivity. Its sign corroborates with that of S at room temperature, confirming that the electron-band dominate the transport properties. R$_H$ shows a weak temperature dependence, similar to that of the less doped samples [20]. Its absolute value at 300 K (-8 10$^{-10}$ m$^3$/C) is much lower than that of the x = 0.23 doping level (-23 10$^{-9}$ m$^3$/C) [20], and it is consistent with the higher electron doping level. The extracted charge carrier density is 7.8 10$^{21}$ cm$^{-3}$, but it should be taken with caution due to the multiband character of the material. With the same reserve, one can calculate the carrier mobility which comes up to $\mu_H = R_H/\rho$ = 5.3 cm$^2$/Vs, a relatively low value for a metal.

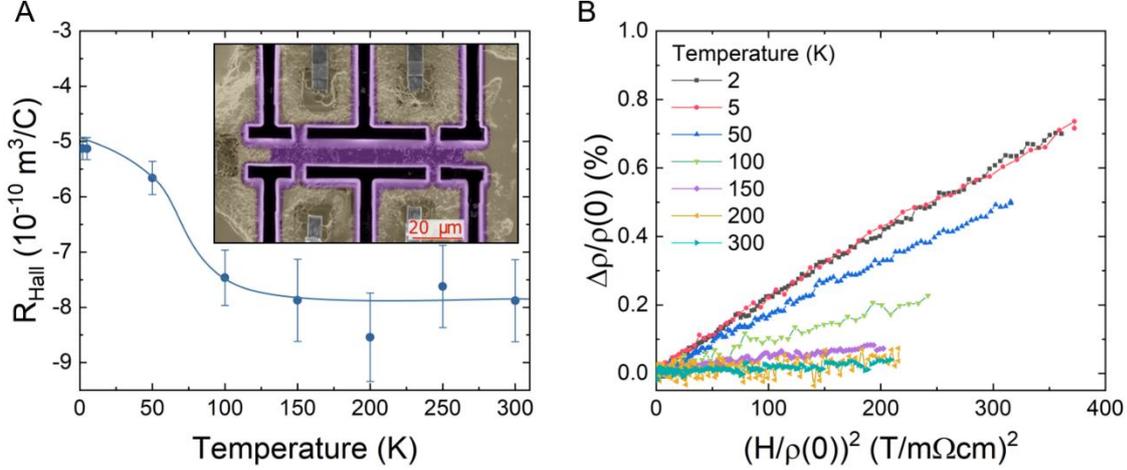

Figure 4. A) Temperature dependence of the Hall coefficient, (the blue line is added as guide for the eyes). In the inset, the SEM picture shows the micro-structured sample. False colors are added to highlight the sample (purple). The sputtered gold layer (yellow) and the FIB-deposited Pt leads (gray) give the electrical contacts. B) Kohler plot of the magnetoresistance plotted at fix temperatures.

The longitudinal magnetoresistance (B || *c*-axis, j in the *ab*-plane) is quadratic in field in the whole temperature range (Fig 4. B) as expected for a metal, especially in the low field limit ($\mu_H B \ll 1$). The data plotted in Kohler's representation shows that they do not fall on a single line, which would be expected for simple metals with a single isotropic relaxation time. It is quite natural that Ca$_{1-x}$La$_x$FeAs$_2$ disobeys the Kohler's law since it has multiple bands taking part in the electronic conduction, with possibly multiple and anisotropic scattering times. The same behaviour is also evident for lower doping levels and in other pnictides [30,31].

**Conclusions**

By growing Ca$_{1-x}$La$_x$FeAs$_2$ (x = 0.56), we have demonstrated that high-pressure synthesis could be the route for exploring the heavily overdoped side of the 112-family phase diagram. Using synchrotron XRD, magnetic susceptibility and transport measurements, the sample was characterized in detail. We have identified a paramagnet-to-antiferromagnet transition around 120 K, well above the highest previously reported transition temperature of 70 K for doping level of x = 0.27. This result suggests a monotonous increase of the antiferromagnetic ordering strength with doping, which is a unique feature among the iron-based superconductors. Such peculiar behaviour as function of doping might be related to the existence of metallic block-layers of As zig-zag chains in the 112-family [10]. The enlarged Fermi surface with doping could enhance the Fermi surface nesting which in turn, increases the antiferromagnetic coupling [32,33].

**Acknowledgments**

We thank Konstantin Semeniuk and Mathieu François Padlewski for useful discussions.

The work at the Laboratory of Physics of Complex Matter was supported by the Swiss National Science Foundation through its SINERGIA network MPBH and Grant No. 200021_175836.

Synchrotron X-ray diffraction measurement was performed at the BM01 station of the Swiss-Norwegian Beam Lines of the European Synchrotron Radiation Facility (ESRF, Grenoble, France).